\documentclass[preprint,pra,12pt,titlepage]{revtex4}%
\usepackage{amsfonts}
\usepackage{amsmath}
\usepackage{amssymb}
\usepackage{graphicx}%
\providecommand{\U}[1]{\protect\rule{.1in}{.1in}}
\begin{document}
\title{Phase bistability and phase bistable patterns in self-oscillatory systems
under a resonant periodic forcing with spatially modulated amplitude}
\author{Germ\'{a}n J. de Valc\'{a}rcel}
\affiliation{Departament d'\`{O}ptica, Universitat de Val\`{e}ncia, Dr. Moliner 50, 46100
Burjassot, Spain, EU}

\begin{abstract}
I consider the problem of self-oscillatory systems undergoing a homogeneous
Hopf bifurcation when they are submitted to an external forcing that is
periodic in time, at a frequency close to the system's natural frequency (1:1
resonance), and whose amplitude is slowly modulated in space. Starting from a
general, unspecified model and making use of standard multiple scales
analysis, I show that the close-to-threshold dynamics of such systems is
universally governed by a generalized, complex Ginzburg-Landau (CGL) equation.
The nature of the generalization depends on the strength and of other features
of forcing: (i) For generic, sufficiently weak forcings the CGL equation
contains an extra, inhomogeneous term proportional to the complex amplitude of
forcing, as in the usual 1:1 resonance with spatially uniform forcing; (ii)
For stronger perturbations, whose amplitude sign alternates across the system,
the CGL equation contains a term proportional to the complex conjugate of the
oscillations envelope, like in the classical 2:1 resonance, responsible for
the emergence of phase bistability and of phase bistable patterns in the
system. Finally I show that case (ii) is retrieved from case (i) in the
appropriate limit so that the latter can be regarded as the generic model for
the close-to-threshold dynamics of the type of systems considered here. The
kind of forcing studied in this work thus represents an alternative to the
classical parametric forcing at twice the natural frequency of oscillations
and opens the way to new forms of pattern formation control in
self-oscillatory systems, what is especially relevant in the case of systems
that are quite insensitive to parametric forcing, such as lasers and other
nonlinear optical cavities.

\end{abstract}
\maketitle

\section{Introduction}

The temporal periodic forcing of spatially extended, self-oscillatory systems
is a classical method to control and excite the formation of spatial patterns
in such systems. This kind of forcing admits a universal description when the
system is operated near the oscillation threshold and forcing acts on an $n:m$
resonance, defined by the relation $\omega_{\mathrm{f}}=\left(  n/m\right)
\left(  \omega_{0}+\delta\omega\right)  $ between the external forcing
frequency $\omega_{\mathrm{f}}$ and the natural frequency of oscillations
$\omega_{0}$, where $n/m$ is an irreducible integer fraction and $\delta
\omega$ is a small mistuning. In such case oscillations settle down in the
system so that any of its variables takes the form $\operatorname{Re}\left[
\kappa u\left(  x,t\right)  e^{i\left(  \omega_{0}+\delta\omega\right)
t}\right]  $, to the leading order, where $\kappa$ is a complex constant and
the slowly varying complex amplitude of the oscillations $u$ verifies the
following generalized complex Ginzburg-Landau equation (CGLE)
\cite{coullet,c2}
\begin{equation}
\partial_{t}u=a_{1}u+a_{2}\nabla^{2}u+a_{3}\left\vert u\right\vert ^{2}%
u+a_{4}\overline{u}^{n-1},\label{cglcoullet}%
\end{equation}
where $\overline{u}$ stands for the complex conjugate of $u$, $a_{i=1,2,3}$
are complex coefficients and $a_{4}$ is proportional to the $m$-th power of
the forcing amplitude. Equation (\ref{cglcoullet}) is valid in principle for
perfectly periodic forcings and has been introduced making use of elegant
symmetry arguments \cite{coullet} as well as has been derived in specific
contexts, such as in chemistry \cite{Walgraef}, making use of standard
multiple scales analysis \cite{Nayfeh}, .

As commented the validity of (\ref{cglcoullet}) requires that the system is
close to the Hopf bifurcation and, additionally that the amplitude of the
external forcing be either constant or slowly varying in time and/or space,
see \cite{German} for the case $n=1$. Here I show that Eq. (\ref{cglcoullet})
with $n=2$ applies as well to systems forced at the 1:1 resonance (hence at
$n=1$) when the forcing amplitude varies on space on a long (but not too long)
scale, to be defined formally below, and the spatial variation involves a sign
alternation of the amplitude. This implies that the system becomes phase
bistable, which is one of the salient features of Eq. (\ref{cglcoullet}) with
$n=2$ \ref{coullet}, in contrast to the phase locking (to just one phase
value) predicted in the case $n=1$. The intuitive picture is that, as the
system is resonantly forced at a 1:1 resonance the phase of the oscillations
tend to lock to a single value that depends in particular on the phase of the
forcing. In our case however forcing displays two opposite phases (two signs)
that are alternating across the system. The system's oscillations have then
two reference phases and can lock to two dynamically equivalent values,
leading to phase bistability. Clearly this must require that the typical
spatial scale over which forcing varies is short as compared with the typical
spatial scale of the unforced system as otherwise the system could lock
locally to the forcing phase at that region. The kind of forcing considered in
this work gets inspiration from another 1:1 resonant forcing, called rocking
\cite{rock}, in which the amplitude of forcing is uniform in space but its
sign alternates along time. Rocking has been considered both theoretically and
experimentally in several systems \cite{r2,r3,r4,r5}.

The analysis presented here is based on the technique of multiple scales and
generalizes the concept of resonant forcing as it considers periodic forcings
in time that are nonuniform across the system on an apparently nonresonant
scale. The derivation allows a rigorous simplified study of both spatially
periodic forcings as well as spatially noisy forcings within a 1:1 resonance,
whenever they verify the conditions imposed below. Generalizations to other
resonances ($n\neq1$ or $m\neq1$), although cumbersome, can be made
straightforwardly following the lines of the present derivation.

The kind of forcing studied here thus represents an alternative to the
classical parametric (or 2:1 resonant) forcing, what is especially relevant in
systems that are insensitive to the latter, like most nonlinear optical
systems. In particular this allows the emergence of phase bistability in the
system (as the term $a_{4}\overline{u}$ breaks the usual continuous phase
symmetry of the classical CGLE down to the discrete one $u\rightarrow-u$
\cite{coullet}) and of phase bistable patterns, like phase domains and phase
domain walls, rolls, hexagons and localized structures.

\section{Model and notation}

We consider a generic one-dimensional system described by $N$ real dynamical
variables $\left\{  U_{i}\left(  x,t\right)  \right\}  _{i=1}^{N}$ whose time
evolution is governed by the following set of real equations written in vector
form,
\begin{equation}
\partial_{t}\mathbf{U}\left(  x,t\right)  =\mathbf{f}\left(  \mu
,\alpha;\mathbf{U}\right)  +\mathcal{D}\left(  \mu,\alpha\right)  \cdot
\nabla^{2}\mathbf{U}, \label{model}%
\end{equation}
where $\mathbf{f}$ is a sufficiently differentiable function of its arguments,
$\nabla^{2}=\partial_{x}^{2}$ is the one-dimensional Laplacian and
$\mathcal{D}\left(  \mu,\alpha\right)  $ is a diffusion matrix. This is the
simplest dependence on derivatives in space-translation invariant systems at
the time it corresponds to\ actual systems of most relevance like,
\textit{e.g.}, reaction-diffusion and nonlinear optical systems. $\mu$ is the
bifurcation parameter and $\alpha\left(  x,t\right)  $ is the forcing
parameter, which is allowed to vary on time and space. Physically $\alpha$ may
represent either an independent parameter, or the modulated part of any other parameter.

We assume that in the absence of forcing ($\alpha=0$) Eq. (\ref{model})
supports a steady, spatially homogeneous state $\mathbf{U}=\mathbf{U}%
_{\mathrm{s}}\left(  \mu\right)  $ $\left(  \partial_{t}\mathbf{U}%
_{\mathrm{s}}=\partial_{x}\mathbf{U}_{\mathrm{s}}=0\right)  $, which looses
stability at $\mu=\mu_{0}$ giving rise to a self-oscillatory, spatially
homogeneous state. In other words, we assume that the reference state
$\mathbf{U}_{\mathrm{s}}$ suffers a homogeneous Hopf bifurcation at $\mu
=\mu_{0}$. We wish to study the small amplitude oscillations that form in the
system close to the bifurcation when the arbitrary parameter $\alpha$ is
periodically modulated in time with a frequency close to that of the free
oscillations and is modulated in space on a long spatial scale to be defined below.

For the sake of convenience we introduce a new vector
\begin{equation}
\mathbf{u}\left(  \mathbf{r},t\right)  =\mathbf{U}\left(  x,t\right)
-\mathbf{U}_{\mathrm{s}}, \label{u}%
\end{equation}
which measures the deviation of the system from the reference state, in terms
of which we rewrite Eq. (\ref{model}) as a Taylor series,
\begin{align}
\partial_{t}\mathbf{u}  &  =\mathbf{F}\left(  \mu,\alpha\right)
+\mathcal{J}\left(  \mu,\alpha\right)  \cdot\mathbf{u}+\mathcal{D}\left(
\mu,\alpha\right)  \cdot\nabla^{2}\mathbf{u}\nonumber\\
&  +\mathbf{K}\left(  \mu,\alpha;\mathbf{u},\mathbf{u}\right)  +\mathbf{L}%
\left(  \mu,\alpha;\mathbf{u},\mathbf{u},\mathbf{u}\right)  +\mathrm{h.o.t.},
\label{modelu}%
\end{align}
where $\mathrm{h.o.t.}$ denotes terms of higher order than $3$ in $\mathbf{u}%
$. As usual, and as we show below, these $\mathrm{h.o.t.}$ have no influence
near the bifurcation whenever it is suprecritical, what we assume.

The different elements of the expansion (\ref{modelu}) are defined as
\begin{align*}
\mathbf{F}\left(  \mu,\alpha\right)   &  =\mathbf{f}_{\mathrm{s}},\\
\mathcal{J}_{ij}\left(  \mu,\alpha\right)   &  =\left[  \partial
f_{i}/\partial U_{j}\right]  _{\mathrm{s}},\\
\mathbf{K}\left(  \mu,\alpha;\mathbf{a},\mathbf{b}\right)   &  =\tfrac{1}{2!}%
%TCIMACRO{\tsum \nolimits_{i,j=1}^{N}}%
%BeginExpansion
{\textstyle\sum\nolimits_{i,j=1}^{N}}
%EndExpansion
\left[  \partial^{2}\mathbf{f}/\partial U_{i}\partial U_{j}\right]
_{\mathrm{s}}a_{i}b_{j},\\
\mathbf{L}\left(  \mu,\alpha;\mathbf{a},\mathbf{b},\mathbf{c}\right)   &
=\tfrac{1}{3!}%
%TCIMACRO{\tsum \nolimits_{i,j,k=1}^{N}}%
%BeginExpansion
{\textstyle\sum\nolimits_{i,j,k=1}^{N}}
%EndExpansion
\left[  \partial^{3}\mathbf{f}/\partial U_{i}\partial U_{j}\partial
U_{k}\right]  _{\mathrm{s}}a_{i}b_{j}c_{k},
\end{align*}
where $\mathbf{a},\mathbf{b},\mathbf{c}$ are arbitrary vectors and the
subscript \textrm{"}$\mathrm{s"}$ denotes $\mathbf{U}=\mathbf{U}_{\mathrm{s}%
}\left(  \mu\right)  $. Vector $\mathbf{F}\left(  \mu,\alpha\right)  $ is
subjected to the condition
\begin{equation}
\mathbf{F}\left(  \mu,0\right)  =0, \label{Fmu0}%
\end{equation}
since in the absence of forcing ($\alpha=0$) the reference state
$\mathbf{u}=0$ is a steady state of Eq. (\ref{modelu}) by hypothesis.
$\mathcal{J}$ is a matrix and vector $\mathbf{K}$ ($\mathbf{L}$) is a
symmetric and bilinear (symmetric and trilinear) form of its two (three) last arguments.

\section{The Hopf bifurcation of the unforced system}

In the absence of forcing ($\alpha=0$) the stability of the reference state
against small perturbations $\delta\mathbf{u}$ is governed by the following
equation
\begin{equation}
\partial_{t}\delta\mathbf{u}=\mathcal{J}\left(  \mu,0\right)  \cdot
\delta\mathbf{u}+\mathcal{D}\left(  \mu,0\right)  \cdot\nabla^{2}%
\delta\mathbf{u,} \label{eqlin}%
\end{equation}
obtained upon linearizing Eq. (\ref{modelu}) for $\alpha=0$ with respect to
$\delta\mathbf{u}$. The general solution to Eq. (\ref{eqlin}) is a
superposition of plane waves of the form
\begin{equation}
\delta\mathbf{u}\left(  \mathbf{r},t\right)  =%
%TCIMACRO{\tsum \nolimits_{j}}%
%BeginExpansion
{\textstyle\sum\nolimits_{j}}
%EndExpansion
\mathbf{w}_{j}\exp\left(  \Lambda_{j}t\right)  \exp\left(  ik_{j}x\right)  ,
\end{equation}
with
\begin{align}
\Lambda_{j}\mathbf{w}_{j}  &  =\mathcal{M}\left(  \mu,k_{j}^{2}\right)
\cdot\mathbf{w}_{j},\label{eigensystem}\\
\mathcal{M}\left(  \mu,k^{2}\right)   &  =\mathcal{J}\left(  \mu,0\right)
-k^{2}\mathcal{D}\left(  \mu,0\right)  , \label{matrixM}%
\end{align}
hence eigenvalues and eigenvectors of matrix $\mathcal{M}$ depend on $k$ only
through $k^{2}$ as $\mathcal{M}$ does (this is a consequence of the assumed
spatial-translation invariance of the unforced system).

As we are assuming that the reference state looses stability at $\mu=\mu_{0}$
via a homogeneous Hopf bifurcation in the absence of forcing, matrix
$\mathcal{M}\left(  \mu,k^{2}\right)  $ must have a pair of complex-conjugate
eigenvalues $\left\{  \Lambda_{1},\Lambda_{2}\right\}  =\left\{
\lambda\left(  \mu,k^{2}\right)  ,\overline{\lambda}\left(  \mu,k^{2}\right)
\right\}  $ (the overbar denotes complex conjugation) governing the
instability, \textit{i.e.}:

\begin{enumerate}
\item[(i)] Close to the bifurcation $\operatorname{Re}\Lambda_{i\geq3}<0$,
while $\operatorname{Re}\lambda$ can become positive for some $k$'s,

\item[(ii)] At the bifurcation $\operatorname{Re}\lambda$ is maximum and null
at $k=0$ (the perturbation with largest growth rate is spatially
homogeneous):
\begin{equation}
\operatorname{Re}\lambda_{0}=0,\left(  \partial_{k}\operatorname{Re}%
\lambda\right)  _{0}=0,\left(  \partial_{k}^{2}\operatorname{Re}%
\lambda\right)  _{0}<0. \label{Relambda0}%
\end{equation}
where, here and in the following,
\end{enumerate}

\begin{center}
$\fbox{ \ subscript $0$ affecting functions denotes $\left\{  \mu=\mu
_{0},\alpha=0,k=0\right\}
\begin{array}
[c]{c}%
.
\end{array}
$ \ }$\\[0pt]
\end{center}

\begin{enumerate}
\item[(iii)] The instability is oscillatory, \textit{i.e.},
\begin{equation}
\operatorname{Im}\lambda_{0}\equiv\omega_{0}\neq0. \label{Imlambda0}%
\end{equation}

\end{enumerate}

Finally, the preceding properties imply that:

\begin{enumerate}
\item[(iv)] All eigenvalues of $\mathcal{M}_{0}=\mathcal{J}_{0}$, see Eq.
(\ref{matrixM}), have negative real part but $\left\{  \lambda_{0}%
,\overline{\lambda}_{0}\right\}  =\left\{  i\omega_{0},-i\omega_{0}\right\}  $.
\end{enumerate}

For the sake of later use we introduce the right and left eigenvectors of
$\mathcal{J}_{0}$ associated with eigenvalues $\left\{  i\omega_{0}%
,-i\omega_{0}\right\}  $,
\begin{equation}%
\begin{array}
[c]{cc}%
\mathcal{J}_{0}\cdot\mathbf{h}=i\omega_{0}\mathbf{h,} & \mathcal{J}_{0}%
\cdot\overline{\mathbf{h}}=-i\omega_{0}\overline{\mathbf{h}},\\
\mathbf{h}^{\dagger}\cdot\mathcal{J}_{0}=i\omega_{0}\mathbf{h}^{\dagger}, &
\mathbf{\,}\overline{\mathbf{h}^{\dagger}}\cdot\mathcal{J}_{0}=-i\omega
_{0}\overline{\mathbf{h}^{\dagger}},
\end{array}
\label{r}%
\end{equation}
where the short-hand notation $\mathbf{h}=\mathbf{w}_{1}\left(  \mu=\mu
_{0},k^{2}=0\right)  ,\overline{\mathbf{h}}=\mathbf{w}_{2}\left(  \mu=\mu
_{0},k^{2}=0\right)  $ has been introduced. These vectors verify the following
orthonormality relations:
\begin{equation}
\mathbf{h}^{\dagger}\cdot\overline{\mathbf{h}}=0,\mathbf{h}^{\dagger}%
\cdot\mathbf{h}=1\text{,}%
\end{equation}
as is triviall to be checked.

\section{Scales}

We are interested in determining the small amplitude solutions that emerge in
the system close to the Hopf bifurcation, a parametric region that we define
by
\begin{equation}
\mu=\mu_{0}+\varepsilon^{2}\mu_{2}, \label{mu}%
\end{equation}
where $\varepsilon$ is a smallness parameter $\left(  0<\varepsilon
\ll1\right)  $. The study is based on the widely used technique of multiple
scales \cite{Nayfeh}. These spatial and temporal scales appear naturally close
to the bifurcation and are those on which the asymptotic dynamics of the
unforced system naturally evolves. As is well known, in a homogeneous Hopf
bifurcation these slow scales are given by
\begin{equation}
T=\varepsilon^{2}t,X=\varepsilon x, \label{scales}%
\end{equation}
which follow from the behaviour of $\lambda$ close to the bifurcation, Eq.
(\ref{mu}), for values of $k$ close to the most unstable mode $k=0$:
\begin{equation}
\lambda\left(  \mu_{0}+\varepsilon^{2}\mu_{2},k^{2}\right)  =\lambda
_{0}+\left(  \partial_{\mu}\lambda\right)  _{0}\varepsilon^{2}\mu_{2}+\frac
{1}{2}\left(  \partial_{k}^{2}\lambda\right)  _{0}k^{2}+\max\left\{
\mathcal{O}\left(  \varepsilon^{4}\right)  ,\mathcal{O}\left(  k^{2}\right)
\right\}  , \label{lambda}%
\end{equation}
where a term $\left(  \partial_{k}\lambda\right)  _{0}k$ has not been included
since $\left(  \partial_{k}\lambda\right)  _{0}=0$ as $\lambda$ is an even
function of $k$. From this equation we obtain, making use of Eq.
(\ref{Relambda0}),
\[
\operatorname{Re}\lambda\left(  \mu_{0}+\varepsilon^{2}\mu_{2},k^{2}\right)
=\left(  \partial_{\mu}\operatorname{Re}\lambda\right)  _{0}\varepsilon^{2}%
\mu_{2}-\frac{1}{2}\left\vert \partial_{k}^{2}\operatorname{Re}\lambda
\right\vert _{0}k^{2}+\max\left\{  \mathcal{O}\left(  \varepsilon^{4}\right)
,\mathcal{O}\left(  k^{2}\right)  \right\}  ,
\]
what indicates that the only modes which can experience linear growth verify
$k=\mathcal{O}\left(  \varepsilon\right)  $, hence the asymptotic dynamics of
the system exhibits spatial variations on a scale $x\sim k^{-1}\sim
\varepsilon^{-1}$ and the slow spatial scale $X=\varepsilon x$ follows. Thus,
setting $k=\varepsilon k_{1}$,
\[
\operatorname{Re}\lambda\left(  \mu_{0}+\varepsilon^{2}\mu_{2},\varepsilon
^{2}k_{1}^{2}\right)  =\varepsilon^{2}\left[  \left(  \partial_{\mu
}\operatorname{Re}\lambda\right)  _{0}\mu_{2}-\frac{1}{2}\left\vert
\partial_{k}^{2}\operatorname{Re}\lambda\right\vert _{0}k_{1}^{2}\right]
+\mathcal{O}\left(  \varepsilon^{4}\right)  ,
\]
which shows that the growth (or decay) of the perturbations occurs on a scale
$t\sim\left(  \operatorname{Re}\lambda\right)  ^{-1}\sim\varepsilon^{-2}$ and
the slow timescale $T=\varepsilon^{2}t$ follows. On the other hand, making use
of Eq. (\ref{Imlambda0}) and setting $k=\varepsilon k_{1}$ again, Eq.
(\ref{lambda}) yields
\[
\operatorname{Im}\lambda\left(  \mu_{0}+\varepsilon^{2}\mu_{2},\varepsilon
^{2}k_{1}^{2}\right)  =\omega_{0}+\varepsilon^{2}\left[  \left(  \partial
_{\mu}\operatorname{Im}\lambda\right)  _{0}\mu_{2}+\frac{1}{2}\left(
\partial_{k}^{2}\operatorname{Im}\lambda\right)  _{0}k_{1}^{2}\right]
+\mathcal{O}\left(  \varepsilon^{4}\right)  ,
\]
whose first term, $\omega_{0}=\mathcal{O}\left(  \varepsilon^{0}\right)  $,
indicates that the original timescale $t$ must be retained, while the rest of
terms do not introduce other relevant timescales.

\subsection{Scales for the forcing}

As for the external forcing $\alpha\left(  x,t\right)  $ we assume that it is
weak, of order $\varepsilon^{2}$, periodic in time at a frequency
$\omega_{\mathrm{f}}$ close to the Hopf frequency $\omega_{0}$, and depends on
space on a scale of order $\varepsilon^{-1/2}$, which is shorter that the
typical spatial scale $X=\varepsilon x$ of the system. We note that this
choice introduces another spatial scale%
\begin{equation}
\xi=\varepsilon^{1/2}x.\label{scalenew}%
\end{equation}
For the sake of definiteness we assume that $\omega_{\mathrm{f}}=\omega
_{0}+\varepsilon^{2}\omega_{2}$ and adopt the following expression for
$\alpha$:
\begin{equation}
\alpha\left(  x,t\right)  =\varepsilon^{2}\alpha_{2}\left(  x,t\right)
=\varepsilon^{2}A\left(  \xi\right)  \exp\left(  i\omega_{2}T\right)
\exp\left(  i\omega_{0}t\right)  +c.c.\label{alpha}%
\end{equation}

We note that the final result of the derivation does not depend on the
harmonic character of the forcing: Any periodic (in fact almost periodic)
$\alpha$ at frequency $\omega_{\mathrm{f}}$ close to $\omega_{0}$ can be
expressed as a Fourier series (with slowly varying coefficients) with
fundamental frequency $\omega_{0}$, of which (\ref{alpha}) is its first term,
and the final result depends only on it \cite{German}. We note for later use
that%
\begin{equation}
\mathbf{F}\left(  \mu,\alpha\right)  =\mathbf{F}\left(  \mu_{0}+\varepsilon
^{2}\mu_{2},\varepsilon^{2}\alpha_{2}\right)  =\varepsilon^{2}\alpha
_{2}\left(  \partial_{\alpha}\mathbf{F}\right)  _{0}+O\left(  \varepsilon
^{4}\right)  ,\label{serF}%
\end{equation}
as follows from (\ref{Fmu0})\textbf{.}

\subsection{Scales for the system's oscillations}

Under all the previous conditions a multiple scale analysis is possible
\cite{Nayfeh} and we look for asymptotic solutions to Eq. (\ref{modelu}) in
the form
\begin{equation}
\mathbf{u}\left(  x,t\right)  =\sum\nolimits_{m=2}^{\infty}\varepsilon
^{m/2}\mathbf{u}_{m/2}\left(  X,T,\xi,t\right)  , \label{expansion}%
\end{equation}
so that the expansion starts at order $\varepsilon$, as usual.

We finally introduce Eqs. (\ref{mu}) and (\ref{alpha}--\ref{expansion}) into
Eq. (\ref{modelu}) making use of the following chain rules for
differentiation
\begin{align}
\partial_{t}\mathbf{u}  &  =\sum\nolimits_{m=2}^{\infty}\varepsilon
^{m/2}\left(  \partial_{t}+\varepsilon^{2}\partial_{T}\right)  \mathbf{u}%
_{m/2},\\
\nabla^{2}\mathbf{u}  &  =\sum\nolimits_{m=2}^{\infty}\varepsilon^{m/2}\left(
\varepsilon\partial_{\xi}^{2}+2\varepsilon^{3/2}\partial_{\xi}\partial
_{X}+\varepsilon^{2}\partial_{X}^{2}\right)  \mathbf{u}_{m/2},
\label{Lapescala}%
\end{align}
and solve at increasing orders in $\varepsilon$. Note that the occurrence of a
term $2\varepsilon^{3/2}\partial_{\xi}\partial_{X}$ in the Laplacian operator
suggests performing the expansion in terms of $\varepsilon^{1/2}$, and not of
$\varepsilon$, see (\ref{expansion}).

\section{The asymptotic analysis}

The general form of Eq. (\ref{modelu}) at any order $\varepsilon^{m}$ is found
to be
\begin{equation}
\mathfrak{J}\left(  \mathbf{u}_{m}\right)  =\mathbf{g}_{m}\left(
X,T,\xi,t\right)  ,\label{orderm}%
\end{equation}
where
\begin{equation}
\mathfrak{J}\left(  \mathbf{u}_{m}\right)  \equiv\partial_{t}\mathbf{u}%
_{m}-\mathcal{J}_{0}\cdot\mathbf{u}_{m},
\end{equation}
and $\mathbf{g}_{m}$ does not depend on $\mathbf{u}_{m}$ but on $\mathbf{u}%
_{n<m}$. Clearly, as $\mathfrak{J}\left(  \exp\left(  i\omega_{0}t\right)
\mathbf{h}\right)  =\mathfrak{J}\left(  \exp\left(  -i\omega_{0}t\right)
\overline{\mathbf{h}}\right)  =0$, see Eq. (\ref{r}), the solvability of Eq.
(\ref{orderm}) requires
\begin{equation}
\frac{\omega_{0}}{2\pi}\int_{t}^{t+2\pi/\omega_{0}}dt^{\prime}\mathbf{h}%
^{\dagger}\cdot\mathbf{g}_{m}\left(  X,T,\xi,t^{\prime}\right)  \exp\left(
-i\omega_{0}t^{\prime}\right)  =0,\label{solvabilitym}%
\end{equation}
(or its equivalent complex-conjugate) which ensures that $\mathbf{g}_{m}$ does
not contain secular terms (proportional to $\exp\left(  i\omega_{0}t\right)
\mathbf{h}$ or to $\exp\left(  -i\omega_{0}t\right)  \overline{\mathbf{h}}$).
Once condition (\ref{solvabilitym}) is verified, the asymptotic solution to
Eq. (\ref{orderm}) reads
\begin{align}
\mathbf{u}_{m}\left(  X,T,\xi,t\right)   &  =u_{m}\left(  X,T,\xi\right)
\exp\left(  i\omega_{0}t\right)  \mathbf{h}+\overline{u}_{m}\left(
X,T,\xi\right)  \exp\left(  -i\omega_{0}t\right)  \overline{\mathbf{h}%
}\nonumber\\
&  +\mathbf{u}_{m}^{\bot}\left(  X,T,\xi,t\right)  ,\label{um}%
\end{align}
where $u_{m}\left(  X,T,\xi\right)  $ is not fixed at this order and the last
term is the particular solution. Note that the solution (\ref{um}) should
involve, in principle, terms proportional to all the eigenvectors of
$\mathfrak{J}\left(  \cdot\right)  $ [which are those of $\mathcal{J}_{0}$,
see Eq. (\ref{eigensystem})]. However all of them (but the first two) are
damped according to $\exp\left[  -\left\vert \operatorname{Re}\Lambda
_{i}\left(  \mu_{0},0\right)  \right\vert \,t\right]  $, since
$\operatorname{Re}\Lambda_{i\geq3}\left(  \mu=\mu_{0},k=0\right)  <0$ (and of
order $\varepsilon^{0}$ by hypothesis), except those associated with $\left(
\mathbf{h},\overline{\mathbf{h}}\right)  $.

\subsection{Order $\varepsilon$}

This is the lowest nontrivial order and%
\begin{equation}
\mathbf{g}_{1}=0,\label{order1}%
\end{equation}
so the solvability condition (\ref{solvabilitym}) at this order is
automatically fulfilled. Then, according to Eq. (\ref{um}),
\begin{equation}
\mathbf{u}_{1}=u_{1}\left(  X,T,\xi\right)  \exp\left(  i\omega_{0}t\right)
\mathbf{h}+\overline{u}_{1}\left(  X,T,\xi\right)  \exp\left(  -i\omega
_{0}t\right)  \overline{\mathbf{h}},\label{u1}%
\end{equation}
where the scaled, slowly varying complex amplitude of oscillations
$u_{1}\left(  X,T,\xi\right)  $ is undetermined at this stage. We need to
continue the analysis in order to meet solvability conditions that fix an
equation for $u_{1}$, which is our goal.

\subsection{Order $\varepsilon^{3/2}$}

At this order we get%
\begin{equation}
\mathbf{g}_{3/2}=0,
\end{equation}
hence the solvability condition is automatically fulfilled again and
$\mathbf{u}_{3/2}$ reads, according to (\ref{um}),%
\begin{equation}
\mathbf{u}_{3/2}=u_{3/2}\left(  X,T,\xi\right)  \exp\left(  i\omega
_{0}t\right)  \mathbf{h}+\overline{u}_{3/2}\left(  X,T,\xi\right)  \exp\left(
-i\omega_{0}t\right)  \overline{\mathbf{h}}. \label{u32}%
\end{equation}

\subsection{Order $\varepsilon^{2}$}

Trivially one has
\begin{equation}
\mathbf{g}_{2}=\alpha_{2}\left(  \partial_{\alpha}\mathbf{F}\right)
_{0}+\mathcal{D}_{0}\cdot\partial_{\xi}^{2}\mathbf{u}_{1}+\mathbf{K}\left(
\mu_{0},0;\mathbf{u}_{1},\mathbf{u}_{1}\right)  . \label{order2}%
\end{equation}
Making use of Eq. (\ref{u1}) and taking into account that $\mathbf{K}$\ is
symmetric and bilinear in its two last arguments, Eq. (\ref{order2}) can be
written as
\begin{align}
\mathbf{g}_{2}  &  =\alpha_{2}\left(  \partial_{\alpha}\mathbf{F}\right)
_{0}+\left(  \mathcal{D}_{0}\cdot\mathbf{h}\right)  \partial_{\xi}^{2}%
u_{1}\left(  X,T,\xi\right)  \exp\left(  i\omega_{0}t\right)  +\left(
\mathcal{D}_{0}\cdot\overline{\mathbf{h}}\right)  \partial_{\xi}^{2}%
\overline{u}_{1}\left(  X,T,\xi\right)  \exp\left(  -i\omega_{0}t\right)
\nonumber\\
&  +2\mathbf{K}\left(  \mu_{0},0;\mathbf{h},\overline{\mathbf{h}}\right)
\left\vert u_{1}\right\vert ^{2}+\mathbf{K}\left(  \mu_{0},0;\mathbf{h}%
,\mathbf{h}\right)  u_{1}^{2}\exp\left(  i2\omega_{0}t\right)  +\mathbf{K}%
\left(  \mu_{0},0;\overline{\mathbf{h}},\overline{\mathbf{h}}\right)
\overline{u}_{1}^{2}\exp\left(  -i2\omega_{0}t\right)  .
\end{align}
The solvability condition (\ref{solvabilitym}) reads now
\begin{equation}
\left(  \mathbf{h}^{\dagger}\cdot\mathcal{D}_{0}\cdot\mathbf{h}\right)
\partial_{\xi}^{2}u_{1}\left(  X,T,\xi\right)  +\left[  \mathbf{h}^{\dagger
}\cdot\left(  \partial_{\alpha}\mathbf{F}\right)  _{0}\right]  \frac
{\omega_{0}}{2\pi}\int_{t}^{t+2\pi/\omega_{0}}dt^{\prime}\alpha_{2}\exp\left(
-i\omega_{0}t^{\prime}\right)  =0,
\end{equation}
which making use of (\ref{alpha}) becomes%
\begin{equation}
c_{2}\partial_{\xi}^{2}u_{1}\left(  X,T,\xi\right)  +c_{4}A\left(  \xi\right)
e^{i\omega_{2}T}=0, \label{solva2}%
\end{equation}
where we defined%
\begin{align}
c_{2}  &  =\mathbf{h}^{\dagger}\cdot\mathcal{D}_{0}\cdot\mathbf{h},\\
c_{4}  &  =\mathbf{h}^{\dagger}\cdot\left(  \partial_{\alpha}\mathbf{F}%
\right)  _{0},
\end{align}

Condition (\ref{solva2}) implies that $u_{1}$ can be expressed as%
\begin{equation}
u_{1}\left(  X,T,\xi\right)  =\left(  c_{4}/c_{2}\right)  u_{A}\left(
\xi\right)  e^{i\omega_{2}T}+U_{1}\left(  X,T\right)  , \label{solu1}%
\end{equation}
where $u_{A}\left(  \xi\right)  $ is the particular solution to%
\begin{equation}
\frac{d^{2}u_{A}\left(  \xi\right)  }{d\xi^{2}}=-A\left(  \xi\right)  ,
\label{uA}%
\end{equation}
and $U_{1}\left(  X,T\right)  $ is a yet undetermined function of the slow
scales $X$ and $T$. Note that the solvability of Eq. (\ref{uA}) requires that
$A$ does not contain a constant term. Hence we assume in the following that%
\begin{equation}
\left\langle A\left(  \xi\right)  \right\rangle =0, \label{promA}%
\end{equation}
where the angular brackets denote averaging over the spatial scale $\xi$.

Finally, according to Eq. (\ref{um}),
\begin{align}
\mathbf{u}_{2}  &  =u_{2}\left(  X,T,\xi\right)  \exp\left(  i\omega
_{0}t\right)  \mathbf{h}+\overline{u}_{2}\left(  X,T,\xi\right)  \exp\left(
-i\omega_{0}t\right)  \overline{\mathbf{h}}\nonumber\\
&  +\mathbf{v}_{0}\left\vert u_{1}\right\vert ^{2}+\mathbf{v}_{2}u_{1}^{2}%
\exp\left(  2i\omega_{0}t\right)  +\overline{\mathbf{v}}_{2}\overline{u}%
_{1}^{2}\exp\left(  -2i\omega_{0}t\right) \nonumber\\
&  +\left[  \mathbf{v}_{A}A\left(  \xi\right)  \exp\left(  i\omega
_{2}T\right)  +\left(  \mathcal{D}_{0}\cdot\mathbf{h}\right)  \partial_{\xi
}^{2}u_{1}\left(  X,T,\xi\right)  \right]  \exp\left(  i\omega_{0}t\right)
\nonumber\\
&  +\left[  \overline{\mathbf{v}}_{A}\overline{A}\left(  \xi\right)
\exp\left(  -i\omega_{2}T\right)  +\left(  \mathcal{D}_{0}\cdot\overline
{\mathbf{h}}\right)  \partial_{\xi}^{2}\overline{u}_{1}\left(  X,T,\xi\right)
\right]  \exp\left(  -i\omega_{0}t\right)  \label{u2}%
\end{align}
where,
\begin{align}
\mathbf{v}_{0}  &  =-2\mathcal{J}_{0}^{-1}\cdot\mathbf{K}\left(  \mu
_{0},0;\mathbf{h},\overline{\mathbf{h}}\right)  ,\label{v0}\\
\mathbf{v}_{2}  &  =-\left(  \mathcal{J}_{0}-i2\omega_{0}\mathcal{I}\right)
^{-1}\cdot\mathbf{K}\left(  \mu_{0},0;\mathbf{h},\mathbf{h}\right)
,\label{v2}\\
\mathbf{v}_{A}  &  =-\left(  \mathcal{J}_{0}-i\omega_{0}\mathcal{I}%
-i\omega_{0}\mathbf{h}\otimes\mathbf{h}^{\dagger}+i\omega_{0}\overline
{\mathbf{h}}\otimes\overline{\mathbf{h}^{\dagger}}\right)  ^{-1}\cdot\left(
\partial_{\alpha}\mathbf{F}\right)  _{0} \label{vA}%
\end{align}
are constant vectors, and $\mathcal{I}$ is the $N\times N$ identity matrix.
Note that both $\mathcal{J}_{0}$ and $\left(  \mathcal{J}_{0}-i2\omega
_{0}\mathcal{I}\right)  $ are invertible since neither $0$ nor $2i\omega_{0}$
are eigenvalues of $\mathcal{J}_{0}$ by hypothesis: otherwise other
eigenvalues different from $\left\{  i\omega_{0},-i\omega_{0}\right\}  $ would
have null real part at the bifurcation. For the same reason $\left(
\mathcal{J}_{0}-i\omega_{0}\mathcal{I}-i\omega_{0}\mathbf{h}\otimes
\mathbf{h}^{\dagger}+i\omega_{0}\overline{\mathbf{h}}\otimes\overline
{\mathbf{h}^{\dagger}}\right)  $ is invertible too since $i\omega_{0}$ is not
an eigenvalue of $\left(  \mathcal{J}_{0}-i\omega_{0}\mathbf{h}\otimes
\mathbf{h}^{\dagger}+i\omega_{0}\overline{\mathbf{h}}\otimes\overline
{\mathbf{h}^{\dagger}}\right)  $ because we removed the subspaces spanned by
$\left\{  \mathbf{h},\overline{\mathbf{h}}\right\}  $.

\subsection{Order $\varepsilon^{5/2}$}

At this order we get%
\begin{equation}
\mathbf{g}_{5/2}=2\partial_{\xi}\partial_{X}\mathbf{u}_{1}+\partial_{\xi}%
^{2}\mathbf{u}_{3/2}+\mathbf{K}\left(  \mu_{0},0;\mathbf{u}_{1},\mathbf{u}%
_{3/2}\right)  =\partial_{\xi}^{2}\mathbf{u}_{3/2}+\mathbf{K}\left(  \mu
_{0},0;\mathbf{u}_{1},\mathbf{u}_{3/2}\right)  ,
\end{equation}
where the last equality comes from (\ref{u1}). The solvability condition
(\ref{solvabilitym}) reduces in this case to%
\[
\int_{t}^{t+2\pi/\omega_{0}}dt^{\prime}\mathbf{h}^{\dagger}\cdot\partial_{\xi
}^{2}\mathbf{u}_{3/2}\exp\left(  -i\omega_{0}t^{\prime}\right)  =0,
\]
as $\mathbf{K}\left(  \mu_{0},0;\mathbf{u}_{1},\mathbf{u}_{3/2}\right)  $ does
not contain terms oscillating as $\exp\left(  \pm i\omega_{0}t\right)  $.
Making use of (\ref{u32}) we get%
\begin{equation}
\partial_{\xi}^{2}u_{3/2}\left(  X,T,\xi\right)  =0.
\end{equation}
Then, either $u_{3/2}=0$ or $u_{3/2}$ does not depend on $\xi$. In the
following we will consider the more general case,%
\begin{equation}
\mathbf{u}_{3/2}=u_{3/2}\left(  X,T\right)  \exp\left(  i\omega_{0}t\right)
\mathbf{h}+\overline{u}_{3/2}\left(  X,T\right)  \exp\left(  -i\omega
_{0}t\right)  \overline{\mathbf{h}}. \label{u32def}%
\end{equation}

Once the solvability condition has been verified, $\mathbf{u}_{5/2}$ can be
written, according to (\ref{um}), as%
\begin{align}
\mathbf{u}_{5/2}  &  =u_{5/2}\left(  X,T,\xi\right)  \exp\left(  i\omega
_{0}t\right)  \mathbf{h}+\overline{u}_{5/2}\left(  X,T,\xi\right)  \exp\left(
-i\omega_{0}t\right)  \overline{\mathbf{h}}\nonumber\\
&  +\mathbf{u}_{5/2}^{\bot}\left(  X,T,\xi,t\right)  ,
\end{align}
where the last term can be computed easily but we will not do as we need not
knowing its expression.

\subsection{Order $\varepsilon^{3}$}

This is the final order to be considered as it provides the shought equation
for the small amplitude of oscillations. At this order we find
\begin{align}
\mathbf{g}_{3} &  =-\partial_{T}\mathbf{u}_{1}+\mu_{2}\left(  \partial_{\mu
}\mathcal{J}\right)  _{0}\cdot\mathbf{u}_{1}\nonumber\\
&  +\mathcal{D}_{0}\cdot\partial_{X}^{2}\mathbf{u}_{1}+2\mathcal{D}_{0}%
\cdot\partial_{X}\partial_{\xi}\mathbf{u}_{3/2}+\mathcal{D}_{0}\cdot
\partial_{\xi}^{2}\mathbf{u}_{2}\nonumber\\
&  +2\mathbf{K}\left(  \mu_{0},0;\mathbf{u}_{1},\mathbf{u}_{2}\right)
+\mathbf{K}\left(  \mu_{0},0;\mathbf{u}_{3/2},\mathbf{u}_{3/2}\right)
+\mathbf{L}\left(  \mu_{0},0;\mathbf{u}_{1},\mathbf{u}_{1},\mathbf{u}%
_{1}\right)  .\label{order3}%
\end{align}
Note in the previous expression that $\partial_{X}\partial_{\xi}%
\mathbf{u}_{3/2}=0$ according to (\ref{u32def}). Application of the
solvability condition (\ref{solvabilitym}) yields, after substituting Eqs.
(\ref{u1}), (\ref{u32def}) and (\ref{u2}) into Eq. (\ref{order3}), making use
of the symmetry and linearity properties of vectors $\mathbf{K}$ and
$\mathbf{L}$, and after simple but tedious algebra,
\begin{equation}
\partial_{T}u_{1}=c_{1}\mu_{2}u_{1}+c_{2}\partial_{X}^{2}u_{1}+c_{3}\left\vert
u_{1}\right\vert ^{2}u_{1}+c_{2}\partial_{\xi}^{2}u_{2}+c_{A}\partial_{\xi
}^{2}A\left(  \xi\right)  e^{i\omega_{2}T}+c_{5}\partial_{\xi}^{4}%
u_{1},\label{cgl1}%
\end{equation}
where
\begin{subequations}
\label{coefficients}%
\begin{align}
c_{1} &  =\mathbf{h}^{\dagger}\cdot\left(  \partial_{\mu}\mathcal{J}\right)
_{0}\cdot\mathbf{h},\label{c1}\\
c_{2} &  =\mathbf{h}^{\dagger}\cdot\mathcal{D}_{0}\cdot\mathbf{h},\label{c2}\\
c_{3} &  =2\mathbf{h}^{\dagger}\cdot\mathbf{K}\left(  \mu_{0},0;\mathbf{h}%
,\mathbf{v}_{0}\right)  +2\mathbf{h^{\dagger}\cdot K}\left(  \mu
_{0},0;\overline{\mathbf{h}},\mathbf{v}_{2}\right)  +3\mathbf{h}^{\dagger
}\cdot\mathbf{L}\left(  \mu_{0},0;\mathbf{h},\mathbf{h},\overline{\mathbf{h}%
}\right)  ,\label{c3}\\
c_{4} &  =\mathbf{h}^{\dagger}\cdot\left(  \partial_{\alpha}\mathbf{F}\right)
_{0},\label{c4}\\
c_{A} &  =\mathbf{h}^{\dagger}\cdot\mathcal{D}_{0}\cdot\mathbf{v}_{A},\\
c_{5} &  =\mathbf{h}^{\dagger}\cdot\mathcal{D}_{0}\cdot\mathcal{D}_{0}%
\cdot\mathbf{h},
\end{align}
are constant coefficients.

\section{The complex Ginzburg-Landau equation}

There remains just substituting (\ref{solu1}) into (\ref{cgl1}). Once this is
done, one can see that in Eq. (\ref{cgl1}) there are terms depending on the
slow spatial scale $X$ as well as terms depending on the "fast" spatial scale
$\xi$ so that Eq. (\ref{cgl1}) can be split into two equations: one containing
just functions of $X$ alone, and one containing functions of $\xi$
(\textit{and} possibly $X$). The terms depending on $\xi$ determine partially
the value of $u_{2}$. The terms depending only on $X$ determine the equation
of motion for $U_{1}$, see Eq. (\ref{solu1}), which is the leading order
amplitude of oscillations.

The result is%
\end{subequations}
\begin{align}
\partial_{T}U_{1} &  =c_{1}\mu_{2}U_{1}+c_{2}\partial_{X}^{2}U_{1}%
+c_{3}\left\vert U_{1}\right\vert ^{2}U_{1}+c_{3}\left(  c_{4}/c_{2}\right)
^{2}\gamma e^{2i\omega_{2}T}\overline{U}_{1}+2c_{3}\left\vert c_{4}%
/c_{2}\right\vert ^{2}\gamma^{\prime}U_{1},\label{cgl2}\\
\gamma &  =\left\langle u_{A}^{2}\left(  \xi\right)  \right\rangle
,\ \ \ \ \gamma^{\prime}=\left\langle \left\vert u_{A}\left(  \xi\right)
\right\vert ^{2}\right\rangle ,\label{gammas}%
\end{align}
where $\left\langle {}\right\rangle $ denotes a spatial average (over the
scale $\xi$) as already introduced. In order to arrive to this equation we
assumed that $\left\langle \left\vert u_{A}\left(  \xi\right)  \right\vert
^{2}u_{A}\left(  \xi\right)  \right\rangle =0$.

It is convenient to remove the explicit time dependence in (\ref{cgl2}) by
performing the following change,%
\begin{equation}
U=U_{1}e^{-i\omega_{2}T},
\end{equation}
and the equation becomes%
\begin{equation}
\partial_{T}U=\left(  c_{1}\mu_{2}+i\omega_{2}\right)  U+c_{2}\partial_{X}%
^{2}U+c_{3}\left\vert U\right\vert ^{2}U+c_{3}\left(  c_{4}/c_{2}\right)
^{2}\gamma\overline{U}+2c_{3}\left\vert c_{4}/c_{2}\right\vert ^{2}%
\gamma^{\prime}U,\label{cgle3}%
\end{equation}
which is of the type (\ref{cglcoullet}) with $n=2$ as anticipated.

Summarizing, for a system like (\ref{model}) close to a homogeneous Hopf
bifurcation, defined as%
\begin{equation}
\mu=\mu_{0}+\varepsilon^{2}\mu_{2},
\end{equation}
and under the type of forcing analyzed along this paper, namely%
\begin{equation}
\alpha\left(  x,t\right)  =\varepsilon^{2}A\left(  \xi\right)  \exp\left(
i\omega_{2}T\right)  \exp\left(  i\omega_{0}t\right)  +c.c.,
\end{equation}
small oscillations emerge in the form%
\begin{equation}
\mathbf{u}_{1}=\varepsilon u_{1}\left(  X,T,\xi\right)  \exp\left(
i\omega_{0}t\right)  \mathbf{h}+\varepsilon\overline{u}_{1}\left(
X,T,\xi\right)  \exp\left(  -i\omega_{0}t\right)  \overline{\mathbf{h}%
}+\mathcal{O}\left(  \varepsilon^{3/2}\right)  ,
\end{equation}
where%
\begin{equation}
u_{1}=\left[  \left(  c_{4}/c_{2}\right)  u_{A}\left(  \xi\right)  +U\left(
X,T\right)  \right]  e^{i\omega_{2}T},
\end{equation}
$u_{A}\left(  \xi\right)  $ just follows passively the forcing via%
\begin{equation}
\frac{d^{2}u_{A}\left(  \xi\right)  }{d\xi^{2}}=-A\left(  \xi\right)  ,
\end{equation}
where the forcing amplitude must verify $\left\langle A\left(  \xi\right)
\right\rangle =0$, and $U$ is governed by (\ref{cgle3}), in which the
different coefficients are defined in (\ref{coefficients}) and in
(\ref{gammas}). Finally the validity of (\ref{cgle3}) requires that
$\left\langle \left\vert u_{A}\left(  \xi\right)  \right\vert ^{2}u_{A}\left(
\xi\right)  \right\rangle =0$. Note that all the conditions imposed on $A$ (or
on $u_{A}$) imply that the sign of $A$ should alternate in space. As an
example, all previous conditions on the forcing are met for the simple case
$A\left(  \xi\right)  =A_{0}\cos\left(  q\xi\right)  $, in which case
$u_{A}\left(  \xi\right)  =\left(  A_{0}/q^{2}\right)  \cos\left(
q\xi\right)  $, and $\gamma=\gamma^{\prime}=\frac{1}{2}\left(  A_{0}%
/q^{2}\right)  ^{2}$.

Finally, Eq. (\ref{cgle3}) is valid whenever $\operatorname{Re}c_{3}\leq0$
(supercritical bifurcation) since otherwise it would lead to unbounded
solutions. If $\operatorname{Re}c_{3}>0$ the bifurcation is subcritical and
the analysis must incorporate higher orders in the $\varepsilon$--expansion.

\section{A remark on the validity of the classical complex Ginzburg-Landau
equation with resonant forcing to the present case}

Equation (\ref{cgle3}) is a CGLE with broken phase invariance, because of the
occurrence of the term proportional to $\overline{U}$. This term is typical of
self-oscillatory systems forced at twice the natural frequency (2:1
resonance), as stated in the Introduction, but here it has been obtained for a
forcing resonant (1:1 resonant in fact) with the oscillations, whose amplitude
is spatially modulated at a "short" spatial scale (shorter than the typical
one in the unforced case). Why has this happened?

Coming back to the 1:1 resonance its universal description is given by the
CGLE (\ref{cglcoullet}) with $n=1$, i.e.%
\begin{equation}
\partial_{t}u=a_{1}u+a_{2}\nabla^{2}u+a_{3}\left\vert u\right\vert ^{2}%
u+a_{4}.\label{forced1}%
\end{equation}
The derivation of this equation requires formally that
\begin{equation}
\alpha=\varepsilon^{3}\alpha_{3}=\varepsilon^{3}A_{3}\left(  X,T\right)
e^{i\omega_{0}t}+c.c.\label{alphaold}%
\end{equation}
of order $\varepsilon^{3}$, not $\varepsilon^{2}$ as we assumed up to here,
and then $a_{4}$ is proportional to $A_{3}\left(  X,T\right)  $, see
\cite{German}. The question is then: Is Eq. (\ref{forced1}) valid even when
forcing is a bit stronger (of order $\varepsilon^{2}$) and acts on shorter
spatial scales, as we are considering in this paper? In order to make a closer
analysis we consider the version of Eq. (\ref{forced1}) adapted to our
notation, as derived in \cite{German},%
\begin{equation}
\partial_{T}u_{1}=c_{1}\mu_{2}u_{1}+c_{2}\partial_{X}^{2}u_{1}+c_{3}\left\vert
u_{1}\right\vert ^{2}u_{1}+c_{4}A_{3}\left(  X,T\right)  ,\label{fG}%
\end{equation}
where all coefficients have the same meaning as in Eq. (\ref{cgle3}).

Assume now that $A_{3}$ is "large" (this does not mean that the actual forcing
$\alpha$ is) and that it varies on a "short" spatial scale $\xi=\varepsilon
^{-1/2}X$, see (\ref{scalenew}) and (\ref{scales}). In particular we consider
formally that%
\begin{equation}
A_{3}=\varepsilon^{-1}A\left(  \varepsilon^{-1/2}X\right)  e^{i\omega_{2}%
T},\label{al3}%
\end{equation}
where the exponential $e^{i\omega_{2}T}$ has been included in order to
consider the same case we have been dealing with, namely that the frequency of
forcing is $\omega_{\mathrm{f}}=\omega_{0}+\varepsilon^{2}\omega_{2}$, see
(\ref{alphaold}) and (\ref{alpha}). Note that with scaling (\ref{al3}) we are
formally in the same situation as in (\ref{alpha}). What we will show next is
that, if in Eq. (\ref{fG}) we consider formally the scaling (\ref{al3}) for
the forcing, even if, apparently, this scaling is at odds with the valifity
conditions applicable to that equation, one obtains the very same Eq.
(\ref{cgle3}) we have obtained before. This means that Eq. (\ref{fG}) can be
regarded as valid even when forcing is "strong" and varies on a "short"
spatial scale.

The derivation is reasonably simple: As a new spatial scale has been
introduced we assume that $u_{1}$ in (\ref{fG}) can be written as%
\begin{equation}
u_{1}\left(  X,T\right)  =u_{1}^{\left(  0\right)  }\left(  X,T,\xi\right)
+\varepsilon^{1/2}u_{1}^{\left(  1/2\right)  }\left(  X,T,\xi\right)
+\varepsilon^{1}u_{1}^{\left(  1\right)  }\left(  X,T,\xi\right)
+\mathcal{O}\left(  \varepsilon^{3/2}\right)  ,
\end{equation}
what implies that the Laplacian will act as $\partial_{X}^{2}\rightarrow
\partial_{X}^{2}+2\varepsilon^{-1/2}\partial_{X}\partial_{\xi}+\varepsilon
^{-1}\partial_{\xi}^{2}$. Then the asymptotic analysis starts. At the leading,
$\varepsilon^{-1}$ order we obtain $\partial_{\xi}^{2}u_{1}^{\left(  0\right)
}=-\left(  c_{4}/c_{2}\right)  A\left(  \xi\right)  e^{i\omega_{2}T}$, what
implies that%
\begin{equation}
u_{1}^{\left(  0\right)  }\left(  X,T,\xi\right)  =\left(  c_{4}/c_{2}\right)
u_{A}\left(  \xi\right)  e^{i\omega_{2}T}+U_{1}\left(  X,T\right)  ,\label{e0}%
\end{equation}
where%
\begin{equation}
\frac{d^{2}u_{A}\left(  \xi\right)  }{d\xi^{2}}=-A\left(  \xi\right)
,\label{dua}%
\end{equation}
which are nothing but Eqs. (\ref{solu1}) and (\ref{uA}), respectively.

The next, $\varepsilon^{-1/2}$, order reads%
\begin{equation}
0=2\partial_{X}\partial_{\xi}u_{1}^{\left(  0\right)  },
\end{equation}
which is identically fulfilled because of (\ref{e0}). Then, at order
$\varepsilon^{0}$ we get%
\begin{equation}
\partial_{T}u_{1}^{\left(  0\right)  }=c_{1}\mu_{2}u_{1}^{\left(  0\right)
}+c_{2}\left(  \partial_{X}^{2}u_{1}^{\left(  0\right)  }+2\partial
_{X}\partial_{\xi}u_{1}^{\left(  1/2\right)  }+\partial_{\xi}^{2}%
u_{1}^{\left(  1\right)  }\right)  +c_{3}\left\vert u_{1}^{\left(  0\right)
}\right\vert ^{2}u_{1}^{\left(  0\right)  }.
\end{equation}

Substitution of (\ref{e0}) leads to%
\begin{align}
&  \partial_{T}U_{1}-c_{1}\mu_{2}U_{1}-c_{2}\partial_{X}^{2}U_{1}%
-c_{3}\left\vert U_{1}\right\vert ^{2}U_{1}-2c_{3}\left\vert c_{4}%
/c_{2}\right\vert ^{2}\gamma^{\prime}U_{1}-c_{3}\left(  c_{4}/c_{2}\right)
^{2}\gamma e^{2i\omega_{2}T}\overline{U}_{1}\nonumber\\
&  =c_{3}\left[  2\left\vert c_{4}/c_{2}\right\vert ^{2}\left(  \left\vert
u_{A}\right\vert ^{2}-\gamma^{\prime}\right)  U_{1}+\left(  c_{4}%
/c_{2}\right)  ^{2}\left(  u_{A}^{2}-\gamma\right)  e^{2i\omega_{2}T}%
\overline{U}_{1}\right]  \nonumber\\
&  +\left(  c_{1}\mu_{2}-i\omega_{2}\right)  \left(  c_{4}/c_{2}\right)
u_{A}e^{i\omega_{2}T}+c_{2}\left(  2\partial_{X}\partial_{\xi}u_{1}^{\left(
1/2\right)  }+\partial_{\xi}^{2}u_{1}^{\left(  1\right)  }\right)  \nonumber\\
&  +c_{3}\left[  \left\vert c_{4}/c_{2}\right\vert ^{2}\left(  c_{4}%
/c_{2}\right)  \left\vert u_{A}\right\vert ^{2}u_{A}e^{i\omega_{2}T}+\left(
\overline{c}_{4}/\overline{c}_{2}\right)  \overline{u}_{A}e^{-i\omega_{2}%
T}U_{1}^{2}+2\left(  c_{4}/c_{2}\right)  u_{A}e^{i\omega_{2}T}\left\vert
U_{1}\right\vert ^{2}\right]  ,\label{long}%
\end{align}
where $\gamma$ and $\gamma^{\prime}$ have been defined as in (\ref{gammas}).
We note that in this equation the left hand side is independent of the short
spatial scale $\xi$. On the other hand no term on the right hand side depends
solely on $X$ (we assume that $\left\langle \left\vert u_{A}\right\vert
^{2}u_{A}\right\rangle =0$ as we have done in the rest of this paper). Then
the solution to (\ref{long}) is simple: equate to zero both sides. Doing it
with the right hand side gives a condition on $2\partial_{X}\partial_{\xi
}u_{1}^{\left(  1/2\right)  }+\partial_{\xi}^{2}u_{1}^{\left(  1\right)  }$,
which we are not interested in. Equating to zero the left hand side yields the
sought equation,
\begin{equation}
\partial_{T}U_{1}=c_{1}\mu_{2}U_{1}+c_{2}\partial_{X}^{2}U_{1}+c_{3}\left\vert
U_{1}\right\vert ^{2}U_{1}+c_{3}\left(  c_{4}/c_{2}\right)  ^{2}\gamma
e^{2i\omega_{2}T}\overline{U}_{1}+2c_{3}\left\vert c_{4}/c_{2}\right\vert
^{2}\gamma^{\prime}U_{1},
\end{equation}
which coincides exactly with (\ref{cgl2}).

This demonstrates that the usual CGLE describing the 1:1 resonant forcing of
self-oscillatory systems, Eq. (\ref{fG}), is valid even when the forcing term
is "large" and varies on a short spatial scale. The reason for this is quite
easy to understand: Equation (\ref{fG}) is valid for positive $\mu_{2}$ (above
the bifurcation), for negative $\mu_{2}$ (below the bifurcation), but even for
$\mu_{2}=0$, as is trivial to be checked. Then one should consider that Eq.
(\ref{fG}) is valid whichever the value of $\mu_{2}$ be (it must be however,
at most, of order $\varepsilon^{0}$). Then a trivial rescaling of (\ref{fG})
with $A_{3}\left(  X,T\right)  =A\left(  X\right)  e^{i\omega_{2}T}$,%
\begin{equation}
\tau=\eta T,\ \ \ \xi=\eta^{1/2}\ X,\ \ \ \psi\left(  \xi,\tau\right)
=\eta^{-1/2}u_{1}\left(  X,T\right)  ,\ \ \ \eta=\mu_{2}\operatorname{Re}%
c_{1},
\end{equation}
leads to%
\begin{equation}
\partial_{\tau}\psi=\left(  1+i\theta\right)  \psi+c_{2}\partial_{\xi}^{2}%
\psi+c_{3}\left\vert \psi\right\vert ^{2}\psi+c_{4}\eta^{-3/2}A_{3}\left(
\eta^{-1/2}\xi\right)  e^{i\eta^{-1}\omega_{2}\tau},
\end{equation}
where $\theta=\frac{\operatorname{Im}c_{1}}{\operatorname{Re}c_{1}}$. Then, we
see that there exists a normalization telling us that, if $\mu_{2}$ is small
(then $\eta$ is too) the forcing term in (\ref{fG}) can be effectively large
and vary on a short spatial scale. In this case the detuning $\omega_{2}$ must
be accordingly small, of order $\eta$, so that $\eta^{-1}\omega_{2}%
=\mathcal{O}\left(  \eta^{0}\right)  $, as otherwise the inhomogeneous term is
highly nonresonant and produces no effect.

\section{Conclusions}

Starting from a general, unspecified model for a spatially extended system
affected by a homogeneous Hopf bifurcation and forced externally by a
perturbation that is resonant in time with the system's oscillations (1:1
resonance) and is spatially modulated (involving sign alternations) at a scale
shorter that the typical one of the unforced system, I have shown that the
close to threshold dynamics of the system is governed by a complex
Ginzburg-Landau equation with a phase symmetry breaking term, proportional to
the complex conjugate of the oscillations amplitude. This term appears in the
universal description of the 2:1 resonance of self-oscillatory systems
(forcing at twice the system's naural frequency) and is responsible for the
emergence of phase bistability and phase bistable patterns, including phase
domain walls, bright solitonic structures and periodic patterns. Thus the kind
of forcing put forward in this paper represents an alternative to the
classical 2:1 periodic forcing.

Finally I have shown that even the classical complex Ginzburg-Landau equation
valid in a 1:1 resonance, which contains an inhomogeneous term (that can vary,
in principle, only on long spatial and time scales), is valid to describe the
phenomenon analyzed here, i.e., in that equation one can consider that the
forcing is "large" and that varies on a "short" spatial scale. Even if not
shown here, the same applies if the forcing term varies on a "short" temporal
scale, a case that has been analyzed in the context of "rocking" \cite{rock},
a 1:1 resonant forcing technique in which the amplitude of forcing is
spatially uniform but varies in time, which has been studied theoretically and
experimentally in different contexts \cite{r2,r3,r4,r5}

I acknowledge fruitful discussions with Kestutis Staliunas and Eugenio
Rold\'{a}n. This work was supported by the Spanish Government and the European
Union FEDER through project FIS2008-06024-C03-01.

\end{document}